\documentclass[runningheads]{llncs}
\usepackage[T1]{fontenc}
\usepackage{graphicx}
\usepackage{url}
\usepackage{subcaption}
\usepackage{multirow}
\begin{document}
\title{Hierarchical Knowledge Graphs for Story Understanding in Visual Narratives}
\author{Yi-Chun Chen\inst{1}\orcidID{0009-0003-4035-9894}}
% \author{Anonymous}

\institute{Yale University, New Haven, CT 06510, USA \\ \email{ychen74@alumni.ncsu.edu} \\ \email{yi-chun.chen@yale.edu}}
% \institute{Yale University, New Haven, CT 06510, USA} 
% % Springer Heidelberg, Tiergartenstr. 17, 69121 Heidelberg, Germany
% \email{ychen74@alumni.ncsu.edu \and yi-chun.chen@yale.edu}

% %
\maketitle              % typeset the header of the contribution
%
% \begin{abstract}
% This paper presents a hierarchical knowledge graph framework for the structured understanding of visual narratives, focusing on multimodal media such as comics. The proposed method decomposes narrative content into multiple levels—from macro-level story arcs to fine-grained event segments—and represents them through integrated knowledge graphs that capture semantic, spatial, and temporal relationships. At the panel level, we construct multimodal graphs that link visual elements such as characters, objects, and actions with corresponding textual components, including dialogue and captions. These graphs are integrated across narrative levels to support reasoning over story structure, character continuity, and event progression.

% We apply our approach to a manually annotated subset of the Manga109 dataset and demonstrate its ability to support symbolic reasoning across diverse narrative tasks, including action retrieval, dialogue tracing, character appearance mapping, and panel timeline reconstruction. Evaluation results show high precision and recall across tasks, validating the coherence and interpretability of the framework. This work contributes a scalable foundation for narrative-based content analysis, interactive storytelling, and multimodal reasoning in visual media.
% \keywords{Hierarchical Graphs \and  Visual Narratives \and Knowledge Graphs \and Comics Analysis \and Narrative Representation}
% \end{abstract}
\begin{abstract}
We present a hierarchical knowledge graph framework for the structured semantic understanding of visual narratives, using comics as a representative domain for multimodal storytelling. The framework organizes narrative content across three levels—panel, event, and macro-event, by integrating symbolic graphs that encode semantic, spatial, and temporal relationships. At the panel level, it models visual elements such as characters, objects, and actions alongside textual components including dialogue and narration. These are systematically connected to higher-level graphs that capture narrative sequences and abstract story structures.

Applied to a manually annotated subset of the Manga109 dataset, the framework supports interpretable symbolic reasoning across four representative tasks: action retrieval, dialogue tracing, character appearance mapping, and timeline reconstruction. Rather than prioritizing predictive performance, the system emphasizes transparency in narrative modeling and enables structured inference aligned with cognitive theories of event segmentation and visual storytelling. This work contributes to explainable narrative analysis and offers a foundation for authoring tools, narrative comprehension systems, and interactive media applications.
\end{abstract}

\section{Introduction}

Visual narratives—such as comics, graphic novels, and manga—are expressive storytelling media that integrate sequential imagery, dialogue, and spatial composition to convey complex events and emotional experiences~\cite{cohn2013visual,mccloud1993understanding}. For computational systems to interpret or generate such narratives, it is essential to account not only for the content within individual panels but also for the evolving relationships between events across time and modality. This necessitates structured representations capable of capturing the hierarchical and multimodal nature of narrative meaning.

While recent advances in multimodal learning and narrative modeling have improved image-text alignment, many existing approaches treat panels as isolated units or focus narrowly on surface-level correspondences~\cite{iyyer2017amazing}. Such models often overlook how narrative meaning emerges across sequential events, particularly when context and structure span multiple representational levels. To address this gap, we introduce a hierarchical graph-based framework that organizes visual narrative content across multiple levels of abstraction. The proposed method integrates visual, textual, and structural features into a unified knowledge graph architecture, enabling localized reasoning at the panel level and aggregated reasoning at broader narrative tiers.

\begin{figure*}[htbp]
    \centering
    \includegraphics[width=0.9\textwidth]{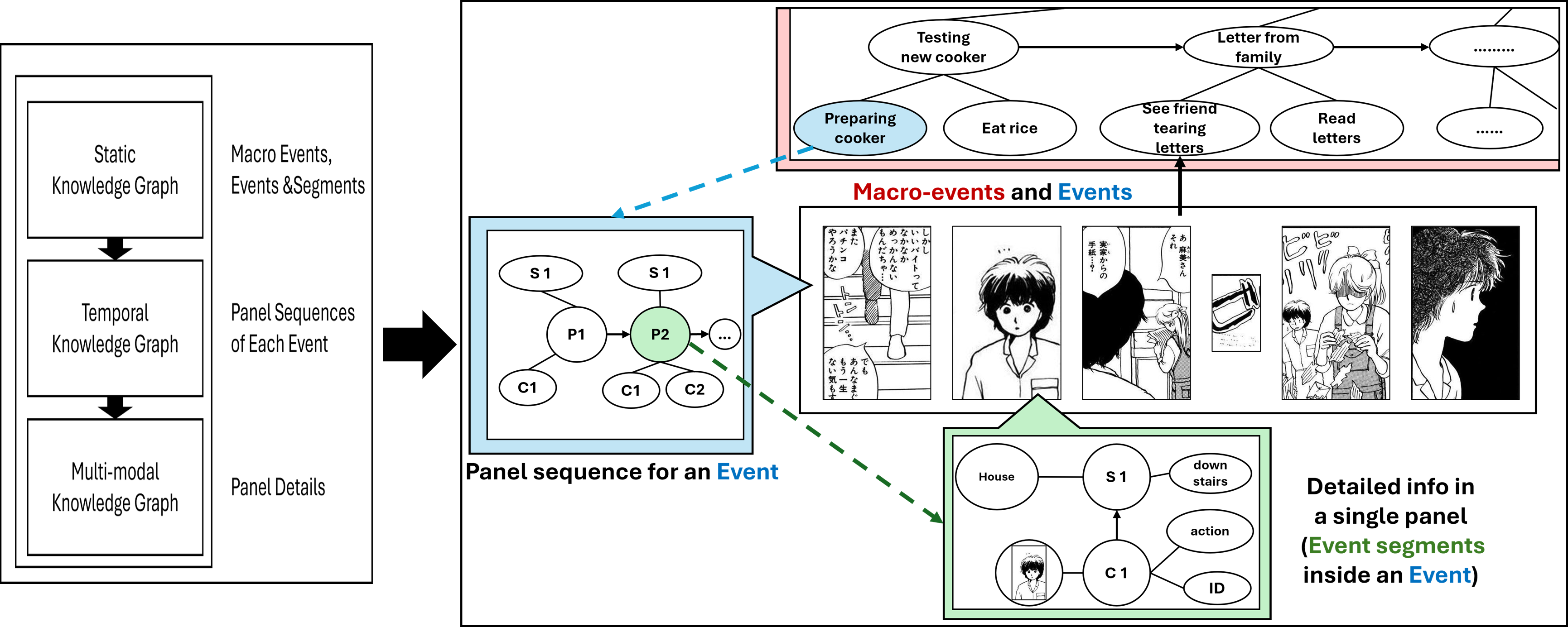}
    \caption{The composition of visual narratives is inherently hierarchical. Story event networks consist of nested events and sub-events, where each event can be broken down into panel sequences. Each panel conveys specific details through multiple modalities, including visual content, text, and spatial composition.}
    \label{fig:basic_structure}
\end{figure*}

Figure~\ref{fig:basic_structure} illustrates the conceptual foundation of our approach, reflecting the inherently layered composition of visual narratives. Individual panels contribute multimodal content that is embedded within event segments, events, and overarching macro-events.

This research is guided by two central questions: (1) How can visual narrative content be systematically represented through a hierarchical structure of knowledge graphs? and (2) How can symbolic reasoning be supported across different levels of narrative abstraction, encompassing static, temporal, and multimodal relationships~\cite{zacks2007eventperception,kim2017deepstory}? To explore these questions, we propose a computational framework that constructs panel-level multimodal graphs and links them to event- and story-level structures capturing semantic and temporal relations.

% The contributions of this work are threefold:
% \begin{itemize}
%     \item A hierarchical framework that integrates multimodal knowledge graphs at multiple narrative levels, supporting compositional and interpretable reasoning from panel-level detail to abstract story structure.
%     \item A graph-based implementation of this framework on a curated and manually annotated visual narrative corpus, demonstrating its ability to capture narrative coherence and multi-level structure.
%     \item Evaluation across four symbolic reasoning tasks, along with qualitative examples showing how the graph structure supports the interpretation of actions, dialogues, character continuity, and event progression.
% \end{itemize}

% This framework offers a scalable and extensible foundation for structured narrative modeling. By bridging symbolic representation and multimodal content, the approach supports applications in automated storytelling, narrative comprehension, and interactive narrative generation.

The contributions of this work are threefold:
\begin{itemize}
    \item A hierarchical framework that integrates multimodal knowledge graphs at multiple narrative levels, supporting compositional and interpretable reasoning from panel-level detail to abstract story structure.
    \item A graph-based implementation of this framework on a curated and manually annotated visual narrative corpus, demonstrating its ability to capture narrative coherence and multi-level structure.
    \item Evaluation across four symbolic reasoning tasks, along with qualitative examples showing how the graph structure supports the interpretation of actions, dialogues, character continuity, and event progression.
\end{itemize}

This framework offers a scalable and extensible foundation for structured narrative modeling. By bridging symbolic representation and multimodal content, the approach supports applications in \textbf{narrative comprehension tasks}, with the potential to inform future work on automated storytelling and interactive narrative generation.

\section{Related Work}
Narrative modeling and visual storytelling have been studied extensively across disciplines, including cognitive science, artificial intelligence, and multimedia computing. This section reviews prior work on event segmentation in narrative understanding, the use of knowledge graphs in storytelling, and computational approaches to visual narratives. These foundations inform our proposed framework, which integrates hierarchical structure and multimodal reasoning via symbolic representations.

\subsection{Narrative Modeling and Story Event Segmentation}

A longstanding focus of cognitive science is understanding how humans segment and comprehend narratives. Early work on story grammars~\cite{thorndyke1977cognitive} proposed that cognitive schemas guide the interpretation of narrative sequences. Later studies~\cite{zwaan1995dimensions,zwaan1996processing,kelter2004representing} introduced dynamic situation models that track shifts in time, space, goals, and characters. The theory of event segmentation~\cite{zacks2007event} further highlights how humans parse continuous experiences into discrete narrative units, a process closely tied to memory formation and prediction~\cite{kurby2008segmentation,sargent2013event}.

These cognitive insights have informed computational approaches to narrative parsing. Biologically inspired models such as cortical reservoirs~\cite{dominey2021narrative}, as well as large language models applied to story segmentation~\cite{michelmann2025large}, aim to mimic human-like processing of narrative structure. However, such models often lack structured and interpretable representations that support fine-grained, multi-level reasoning. This motivates the use of hierarchical knowledge graphs that can explicitly encode narrative segmentation and enable cross-level inference, as pursued in our work.

Beyond visual media, substantial effort has gone into text-based narrative annotation. Finlayson and colleagues developed frameworks for representing story functions and roles, including automatic inference of Proppian functions in folktales~\cite{finlayson2016inferring}, and created the Story Workbench as a tool for authoring and analyzing narrative corpora~\cite{finlayson2011story}. These approaches emphasize event segmentation, role labeling, and semantic coherence in textual stories. 

Our framework shares this goal of structuring narrative meaning but extends it to visual narratives, where multimodal elements such as camera shots, panel composition, and visual–textual alignment must be represented alongside event structure.

\subsection{Knowledge Graphs in Storytelling and Multimodal Media}

Knowledge graphs have emerged as a powerful representation for modeling structured relationships across static, dynamic, and multimodal narrative domains. Surveys by Liang et al.~\cite{liang2024survey} and Lymperaiou and Stamou~\cite{lymperaiou2024survey} highlight how knowledge graphs support the encoding of entities, events, and their relationships in interpretable formats. These structures enable a variety of downstream tasks, including event prediction, narrative generation, and character modeling.

Prior work has explored data-to-text storytelling using graph-based planning~\cite{mishra2019storytelling} and multimedia story generation through linked data and neural models~\cite{renzi2023storytelling}. Hierarchical narrative structures have also been addressed~\cite{akimoto2017computational,lee2020learning}, although not always through explicit graph representations. More recent approaches have incorporated visual input and graph-based reasoning for grounded storytelling~\cite{hsu2020knowledge,xu2021imagine}, indicating the growing interest in integrating relational and perceptual knowledge. In the domain of interactive narrative authoring, recent ICIDS research has explored the use of knowledge graphs to support authoring workflows and mixed-initiative storytelling. Abhilash and Nack~\cite{abhilash2024design} propose design principles for constructing customized knowledge graphs for interactive digital narrative tools. Knight and Eladhari~\cite{knight2024mixed} investigate the creation of comics in mixed initiative as part of artistic practice, further underscoring the growing relevance of symbolic representations in digital narrative systems.

Despite these advances, most existing methods focus on generation rather than comprehension, and relatively few provide fine-grained, interpretable representations. In addition, hierarchical reasoning across levels of abstraction, such as panels, events, and story arcs, remains underexplored. Our framework directly addresses this gap by linking panel-level, event-level, and temporal graphs to support symbolic, cross-modal narrative reasoning.

\subsection{Computational Approaches to Visual Narratives}

Comprehending visual narratives requires integrating insights from cognitive psychology and computational modeling. McNamara and Magliano~\cite{mcnamara2009toward} proposed a cognitive model centered on inference making, coherence construction, and memory integration—key mechanisms in event segmentation. In comics research, Cohn's Visual Narrative Grammar (VNG)~\cite{cohn2014architecture,cohn2015getting,cohn2020visual,cohn2021starring} formalizes the roles and structures of panel sequences, providing a hierarchical theory of visual storytelling that has shaped both cognitive and computational studies. The Scene Perception and Event Comprehension Theory (SPECT)~\cite{loschky2018viewing,loschky2020scene} extends these ideas by linking perceptual processing with narrative structure, emphasizing attention and temporal organization. Building on this, Martens et al.~\cite{martens2020visual} introduced the Visual Narrative Engine, which operationalizes narrative structure using scene and character-level features.

Machine learning approaches have explored narrative alignment and inference across panels. Iyyer et al.~\cite{iyyer2017amazing} applied deep learning to model panel transitions, while Chen and Jhala~\cite{chen2021computational} proposed a manga-specific framework highlighting modality-specific cues. Recent cognitive-inspired work has examined event model updating during reading~\cite{brich2024construction}, and hierarchical panel-to-event models have been developed to bridge visual perception with narrative structure~\cite{chen2023framework}. Although our evaluation focuses on static comics, the proposed framework is intended as a symbolic scaffold that can extend to interactive digital storytelling. Hierarchical graph structures can support branching narratives, dynamic querying, and mixed-initiative authoring, linking multimodal comprehension with generative and interactive systems~\cite{riedl2010narrative,young2013plans}. In this sense, our work complements prior interactive narrative research by emphasizing the representational layer that underpins both interpretability and future integration into author-facing tools.

\setlength{\parskip}{0.5\baselineskip}
These studies confirm that visual narratives are inherently hierarchical and multimodal, spanning sequential panels, temporally organized events, and overarching narrative arcs. However, few prior works offer interpretable symbolic structures that support reasoning across these levels. Our framework addresses this need by constructing and linking multimodal knowledge graphs that operate across narrative tiers to support structured and scalable story comprehension.

\section{Methodology}

This framework models visual narratives using a hierarchy of knowledge graphs that represent entities, actions, and relationships across multiple levels of abstraction. This section outlines the construction of panel-level multimodal graphs, sequence-level temporal graphs, and event-level semantic graphs, as well as their integration into a unified structure. We also describe the hierarchical annotation schema used to support this modeling and the symbolic reasoning functions enabled by the resulting graph representations.

\subsection{Dataset and Annotations}

We construct the hierarchical knowledge graphs on a manually annotated subset of the publicly available Manga109 corpus~\cite{fujimoto2016manga109}. 
Specifically, we annotated two complete story arcs: \textit{Aisazu Niha Irarenai} (Romance) and \textit{Akkera Kanjinchou} (Battle). These arcs were selected for their narrative coherence and genre diversity, allowing evaluation across both introspective romance and action-oriented battle genres. The combined annotations span 111 panels, 102 event segments, 39 events, and 12 macro-events, providing multi-tier coverage suitable for symbolic reasoning.

Each page was segmented into individual panels and annotated across multiple dimensions, including visual content, textual elements, camera shot types, and narrative structure. Annotation was performed manually by the authors and verified through repeated passes. Each panel was annotated for characters, actions, dialogue, camera shots, and event segmentation across three narrative levels. Annotations are hierarchically organized to reflect panel-level descriptions, mid-level events, and high-level macro-events. Additional labels capture character actions, environmental objects, dialogue, and event groupings. These structured annotations form the basis for constructing multimodal and temporal knowledge graphs across three tiers of narrative abstraction. Although the current study focuses on two arcs, the annotation schema and hierarchical graph design are general and can be extended to longer and more diverse visual narratives.

% \subsection{Hierarchical Graph Design}

% To support symbolic reasoning across different levels of narrative abstraction, we propose a three-tier knowledge graph framework that reflects the compositional structure of visual narratives. It is important to note that the hierarchical knowledge graph described here constitutes a conceptual framework: the schema design, cross-level relations, and reasoning principles are independent of any specific implementation. The framework includes:
\subsection{Hierarchical Graph Design}

To enable symbolic reasoning across multiple levels of narrative abstraction, we propose a three-tier knowledge graph framework that mirrors the compositional structure of visual narratives.  The hierarchical design should be understood as a conceptual framework: it specifies the schema, cross-level relations, and reasoning principles that organize narrative information, while remaining independent of any particular software implementation. This distinction allows the same schema to be realized through different technical pipelines, ranging from lightweight traversal code to hybrid neural–symbolic methods. The framework includes:

\begin{itemize}
  \item \textbf{Panel-level multimodal graphs}, encoding visual-textual content and intra-panel semantics;
  \item \textbf{Sequence-level temporal graphs}, modeling the reading and narrative order of events;
  \item \textbf{Event-level semantic graphs}, representing abstract narrative structure across multiple pages.
\end{itemize}

Figure~\ref{fig:basic_structure} (see Section~\ref{fig:basic_structure}) illustrates this layered architecture and its alignment with visual storytelling principles.

\paragraph{Panel-Level: Multimodal Knowledge Graphs}  
Each panel is represented as a graph of visual entities (characters, objects, backgrounds) and textual components (dialogue, captions). Edges encode semantic roles, actions, and spatial or referential links. This level supports localized reasoning tasks such as speaker identification, image-text alignment, and symbolic content extraction. The structure is designed to interface with vision-language models for future integration.

\paragraph{Sequence-Level: Temporal Knowledge Graphs}  
The temporal graph connects panels and event segments into a directed acyclic graph, capturing reading order and narrative progression. It distinguishes visual chronology from narrative causality, accounting for nonlinear patterns such as flashbacks or ellipses. Nodes represent either panels or event segments, and edges denote temporal relations such as \textit{precedes}. This layer supports timeline reconstruction and temporal inference.

\paragraph{Event-Level: Semantic Knowledge Graphs}  
The top-level semantic graph abstracts the narrative into macro-events, events, and sub-events linked by relations such as \textit{subevent-of}, \textit{precedes}, or \textit{co-occurs}. This level enables high-level reasoning, including summarization, structural comparison, and planning for generative storytelling systems.

\subsection{Hierarchical Annotation Design}

To model the temporal and semantic organization of visual narratives, we define a three-level event hierarchy grounded in cognitive theories of visual storytelling, especially Cohn’s Visual Narrative Grammar~\cite{cohn2013visual}. This hierarchy forms the conceptual scaffold for graph construction.

\textbf{Macro-events} are the most abstract narrative units, corresponding to major arcs that span multiple pages. They align with key themes or structural transitions in the story.

\textbf{Events} represent mid-level narrative situations embedded within macro-events. These units are defined by shifts in character goals, interactions, or settings, typically unfolding over multiple panels.

\textbf{Event segments} are fine-grained narrative units associated with short panel sequences. They encode moment-to-moment changes, reactions, or micro-actions, often instantiating narrative roles such as Establishers, Peaks, or Releases.

\begin{figure}[t]
  \centering
  \includegraphics[width=0.8\linewidth]{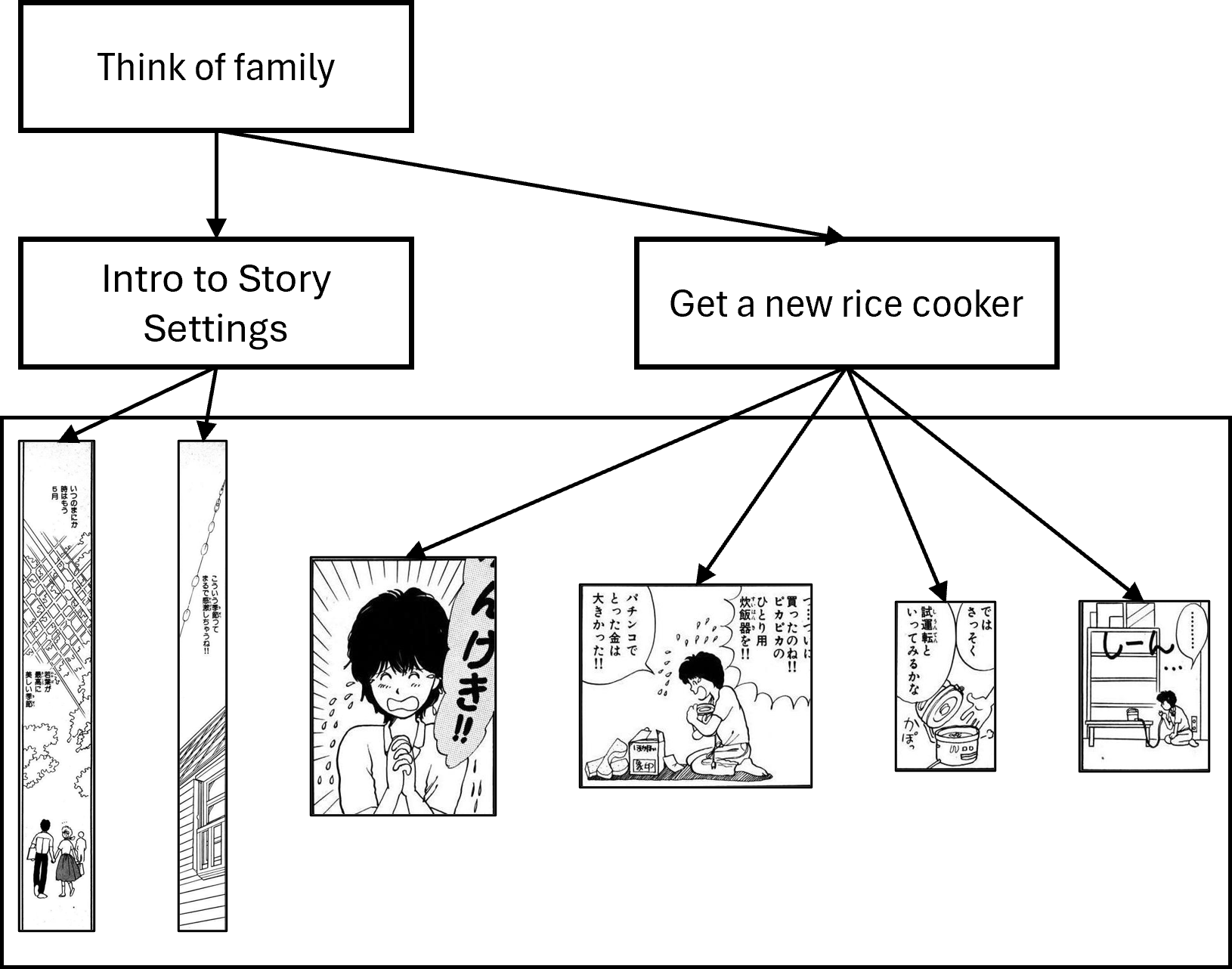}
  \caption{Hierarchical narrative event design. Panels are grouped into event segments, which instantiate mid-level events and aggregate into macro-events.}
  \label{fig:event-hierarchy}
\end{figure}

Figure~\ref{fig:event-hierarchy} illustrates this three-tier structure, showing how panels are grouped into segments, which instantiate events and roll up into macro-events. This organization enables reasoning across narrative scales.

\subsubsection{Annotation Schema: Visual and Textual Elements}

To support multimodal narrative modeling, we define an annotation schema capturing key visual and textual elements within each panel. These annotations serve as the input for constructing panel-level graphs and linking to higher-level structures.

\paragraph{Visual Modality}  
Visual annotations describe entities, spatial layout, and camera composition:

\begin{itemize}
  \item \textbf{Panel Visual}: Path to the panel image and, optionally, an image embedding for integration with visual models.
  \item \textbf{Camera Shot Type}: Annotated using cinematographic terms (e.g., Long Shot, Medium Shot, Close Shot) to capture framing and narrative function.
  \item \textbf{Visual Entities}: Characters, objects, and actions are annotated, with action-agent links supporting structured relation extraction.
\end{itemize}

\paragraph{Textual Modality}  
Textual elements convey narrative content:

\begin{itemize}
  \item \textbf{Dialogue}: Text in speech balloons is linked to speaker entities for tracking conversations and emotional tone.
  \item \textbf{Captions}: Narration and descriptive text provide context, especially for implicit visual or temporal information.
  \item \textbf{Event Descriptions}: Natural language summaries attached to segments or events support semantic abstraction and graph construction.
\end{itemize}

Together, these annotations form a multimodal semantic layer grounded in both surface-level and structural narrative features.

\subsection{Knowledge Graph Integration}

To enable cross-level reasoning, we integrate the panel-, sequence-, and event-level graphs into a single hierarchical representation. Edges such as \textit{instantiates} (from panels to segments) and \textit{subevent-of} (from segments to events and events to macro-events) establish structural links between levels.

Symbolic references such as \textit{refers-to} preserve entity identity across levels, supporting queries like: “Which characters act within this macro-event?” or “Which panels instantiate the final segment of an event?” This unified structure enables bottom-up inference and top-down structural querying across the narrative.

\subsection{Implementation}
\label{sec:integration}

The graph construction pipeline is implemented in Python using the \texttt{NetworkX} library. Annotations and graphs are stored in an interoperable JSON format. The modular system supports incremental construction, editing, and visualization at each narrative level.

All tools, including annotation utilities, graph schemas, and reasoning scripts, will be released publicly upon acceptance to support reproducibility and further research.

\section{Experiments}

We evaluate the utility of our hierarchical knowledge graph framework through a set of symbolic reasoning tasks that reflect core narrative comprehension capabilities. Rather than optimizing for predictive accuracy or large-scale generalization, the goal of these experiments is to demonstrate how symbolic traversal over structured representations can support interpretable, cross-level story understanding in comics.

Specifically, we assess the framework's ability to handle multi-tier reasoning over multimodal content using four targeted narrative tasks: action retrieval, dialogue reconstruction, character appearance mapping, and panel sequence alignment. These tasks highlight the expressiveness and compositionality of the unified graph structure in enabling narrative inference.

\subsection{Dataset and Setup}

All experiments are conducted using a manually annotated subset of the Manga109 dataset~\cite{fujimoto2016manga109}, a professionally curated corpus of Japanese comics exhibiting diverse artistic and narrative structures. Each panel in the selected story is annotated with multimodal information, including characters, objects, camera shot types, dialogue, captions, and hierarchical narrative labels spanning macro-events, events, and event segments.

Using these annotations, we construct knowledge graphs at three levels: panel-level multimodal graphs, sequence-level temporal graphs, and event-level semantic graphs. The graphs are then integrated into a unified representation through cross-level relations such as \textit{instantiates} and \textit{subevent-of}, as described in Section~\ref{sec:integration}. This multi-level graph serves as the basis for all symbolic reasoning functions.

\subsection{Reasoning Tasks}

We define four narrative reasoning tasks to evaluate the framework’s support for structured inference across abstraction levels:

\textbf{1. Action Retrieval by Macro-event.}  
Given a macro-event, the system identifies all associated actions. Reasoning involves traversing from macro-event nodes to their constituent events, segments, and panels, then extracting \texttt{action} nodes from the visual annotations.

\textbf{2. Dialogue Trace by Event.}  
This task evaluates the system’s ability to reconstruct event-specific dialogue. It tests the alignment between events and panels, and the association between panels and textual components such as speech balloons.

\textbf{3. Character Appearance Mapping.}  
The system maps each character to the panels in which they appear, supporting continuity tracking and reasoning about character presence throughout the narrative.

\textbf{4. Panel Timeline Reconstruction.}  
This task evaluates whether the system can recover the correct panel reading order within a macro-event, validating the temporal alignment encoded in the sequence-level graph.

\subsection{Results and Qualitative Examples}

We apply each symbolic reasoning function to the unified graph and extract outputs from the \textit{Think of family} macro-event and the \textit{Intro\_1} event. These examples demonstrate the framework’s capacity for interpretable and coherent narrative inference.

\textbf{Actions in macro-event \textit{Think of family}:} \\  
\indent \texttt{hold\_hand}, \texttt{look\_at\_letter}, \texttt{cook\_rice}, \texttt{walk\_away}

\textbf{Dialogues in event \textit{Intro\_1}:}
\begin{itemize}
    \item ``Before I knew it, it was May, the season when young leaves are the most beautiful.''
    \item ``I had just started living on my own.''
\end{itemize}

\textbf{Character appearances:}
\begin{itemize}
    \item \texttt{A}: appears in panels \texttt{[0\_0\_0, 0\_0\_1, 0\_1\_1, ...]}
    \item \texttt{B}: appears in panels \texttt{[0\_0\_1, 0\_1\_0, ...]}
\end{itemize}

\textbf{Panel sequence in \textit{Think of family}:}  
\texttt{[0\_0\_0, 0\_0\_1, 0\_0\_2, ..., 0\_2\_2]}

These examples illustrate how symbolic traversal over a structured graph can extract high-level narrative elements from multimodal content.

\subsection{Evaluation Methodology and Metrics}

We adopt a hybrid evaluation approach combining qualitative interpretation with lightweight quantitative analysis to assess the framework’s symbolic reasoning capabilities.

\begin{itemize}
    \item \textbf{Qualitative Analysis:} We manually examine outputs to assess their coherence, plausibility, and alignment with expected narrative logic. This includes verifying whether inferred actions, dialogues, and sequences make sense in the story context.

    \item \textbf{Quantitative Metrics:}
    \begin{itemize}
        \item \textbf{Entity Coverage:} Measures the proportion of ground-truth characters or actions successfully retrieved from the graph. This applies to Tasks 1 and 3.
        \item \textbf{Dialogue Recall:} Computes the percentage of annotated dialogue recovered through symbolic traversal, used in Task 2.
        \item \textbf{Ordering Accuracy:} Evaluates agreement between reconstructed panel order and annotated reading sequence, used in Task 4.
    \end{itemize}
\end{itemize}

This evaluation emphasizes interpretability over predictive performance, demonstrating that symbolic reasoning over narrative graphs provides a viable foundation for structured story understanding.

\subsection{Reasoning Functions}

We implement four symbolic graph traversal functions aligned with the narrative elements evaluated in each task:

\textbf{Action Retrieval by Macro-event.}  
Starting at a macro-event node, we traverse through its events and segments to reach panel nodes, and extract actions via \textit{has\_action} relations. This function enables reasoning over visual annotations within a semantic structure.

\textbf{Dialogue Trace by Event.}  
We trace dialogue by navigating from events to panel-level textual nodes via \textit{has\_textual}, \textit{part\_of}, and \textit{content\_of} relations. This supports event-aligned dialogue reconstruction.

\textbf{Character Appearance Mapping.}  
Character nodes are identified via \textit{has\_character} relations from panel visual nodes. Reverse traversal links each character to the panels in which they appear.

\textbf{Panel Timeline by Macro-event.}  
The macro-event is traversed to collect associated panels, which are then sorted based on annotated reading order to reconstruct the narrative timeline.

These functions illustrate the use of symbolic queries over hierarchical narrative graphs for interpretable visual story comprehension.

\subsection{Ground-Truth Construction}

We construct task-specific gold-standard references using annotated data to evaluate symbolic reasoning results:

\textbf{Task 1: Action Retrieval.}  
We extract subject–verb–object tuples from annotated panels and isolate the verb as the action label. All actions associated with a macro-event are deduplicated to form the ground-truth set.

\textbf{Task 2: Dialogue Trace.}  
Annotated dialogue is grouped by event, with duplicates removed to create reference sets for comparison.

\textbf{Task 3: Character Appearance.}  
Character occurrences are grouped by panel and event to define ground-truth mappings.

\textbf{Task 4: Panel Timeline Reconstruction.}  
Panels are ordered according to annotated reading sequences within each macro-event to serve as reference timelines.

These ground-truth sets enable consistent evaluation of symbolic graph reasoning across multiple narrative dimensions.

\section{Result}

This section presents both qualitative and quantitative results from applying our hierarchical knowledge graph framework to visual narrative reasoning. We evaluate the framework’s interpretability and symbolic inference capacity through visual examples and structured reasoning tasks, demonstrating its ability to support high-level narrative understanding.

We evaluate the framework through four structured reasoning tasks: 
(1) \textbf{action retrieval}, identifying character actions within panels or events; 
(2) \textbf{dialogue tracing}, attributing utterances to characters and events; 
(3) \textbf{character appearance mapping}, locating where characters participate across segments or events; and 
(4) \textbf{timeline reconstruction}, ordering panels and events according to narrative flow. 
These tasks are directly grounded in the annotated graph structure and serve as interpretable probes of narrative comprehension. Unlike predictive benchmarks that emphasize generalization accuracy, our evaluation is interpretability-oriented: the tasks highlight how symbolic relations reconstruct narrative content and expose reasoning paths that are transparent to human inspection.

\subsection{Qualitative Examples}

To illustrate the representational capacity of the hierarchical knowledge graph, we present examples at the panel, sequence, and event levels. These visualizations highlight how multimodal entities, temporal order, and abstract narrative structure are encoded symbolically across layers, enabling interpretable reasoning and content reconstruction.

\begin{figure}[t]
  \centering
  \begin{subfigure}[t]{0.3\linewidth}
    \centering
    \includegraphics[width=\linewidth]{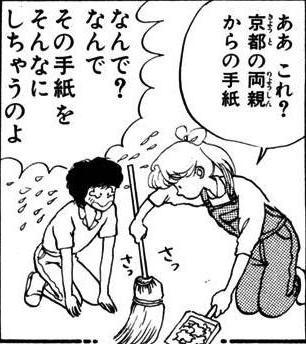}
    \caption{Input panel from \textit{Aisazu Niha Irarenai}.}
    \label{fig:panel-image}
  \end{subfigure}
  \hfill
  \begin{subfigure}[t]{0.6\linewidth}
    \centering
    \includegraphics[width=\linewidth]{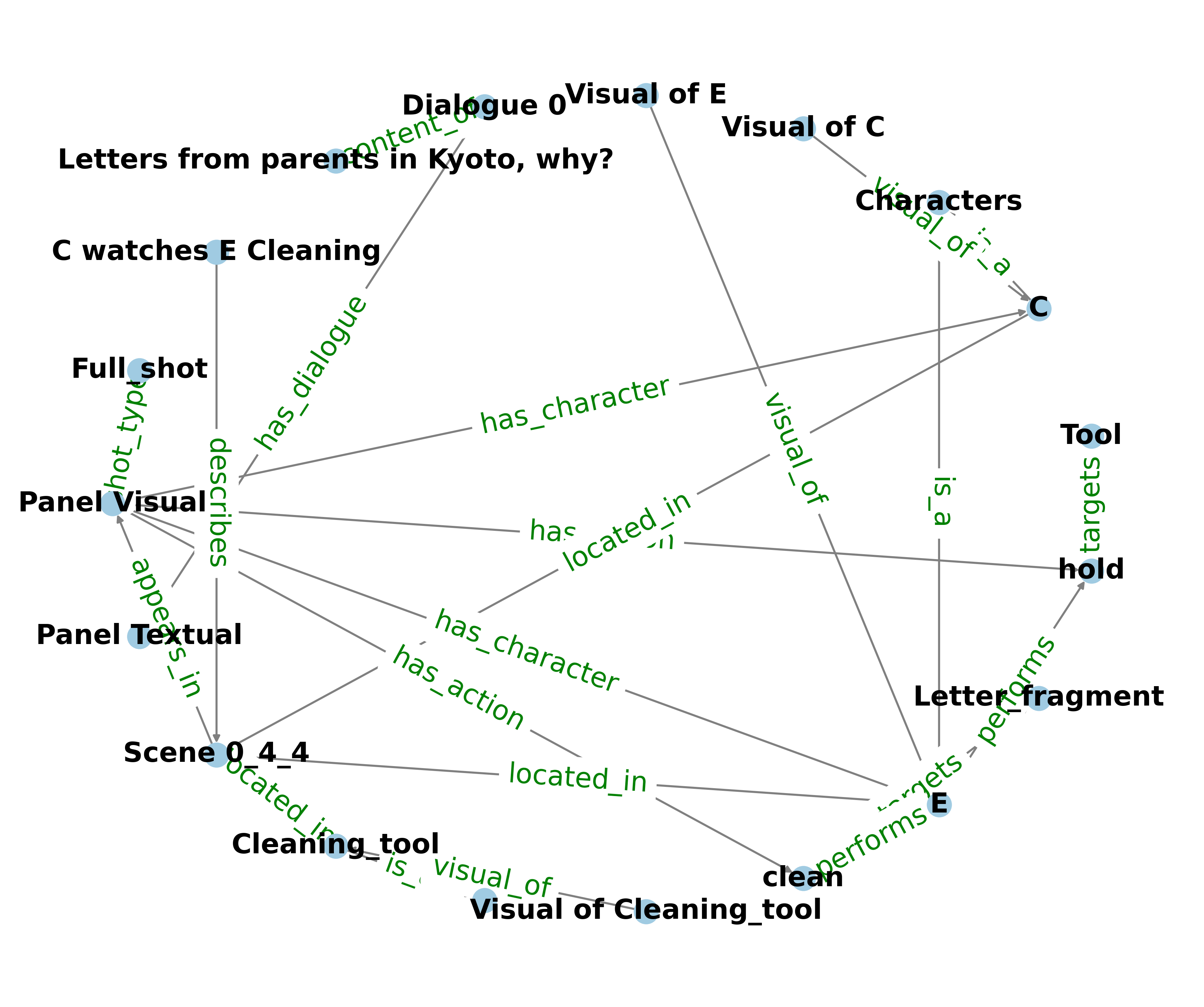}
    \caption{Corresponding panel-level knowledge graph.}
    \label{fig:panel-graph}
  \end{subfigure}
  \caption{Example of panel-level representation. (a) shows the original panel, and (b) shows its multimodal knowledge graph encoding visual and textual entities with semantic links.}
  \label{fig:panel-kg}
\end{figure}

Figure~\ref{fig:panel-kg} shows a worked example from \textit{Aisazu Niha Irarenai}. Subfigure~(a) depicts the original panel where two characters exchange dialogue, and subfigure~(b) shows the corresponding panel-level graph. Nodes represent characters, actions, and dialogue, while edges encode semantic roles and discourse associations. From this representation, a structured query can retrieve utterances associated with each character, demonstrating how localized reasoning tasks such as speech attribution and action recognition are supported.

\begin{figure}[t]
  \centering
  \begin{subfigure}[t]{0.3\linewidth}
    \centering
    \includegraphics[width=\linewidth]{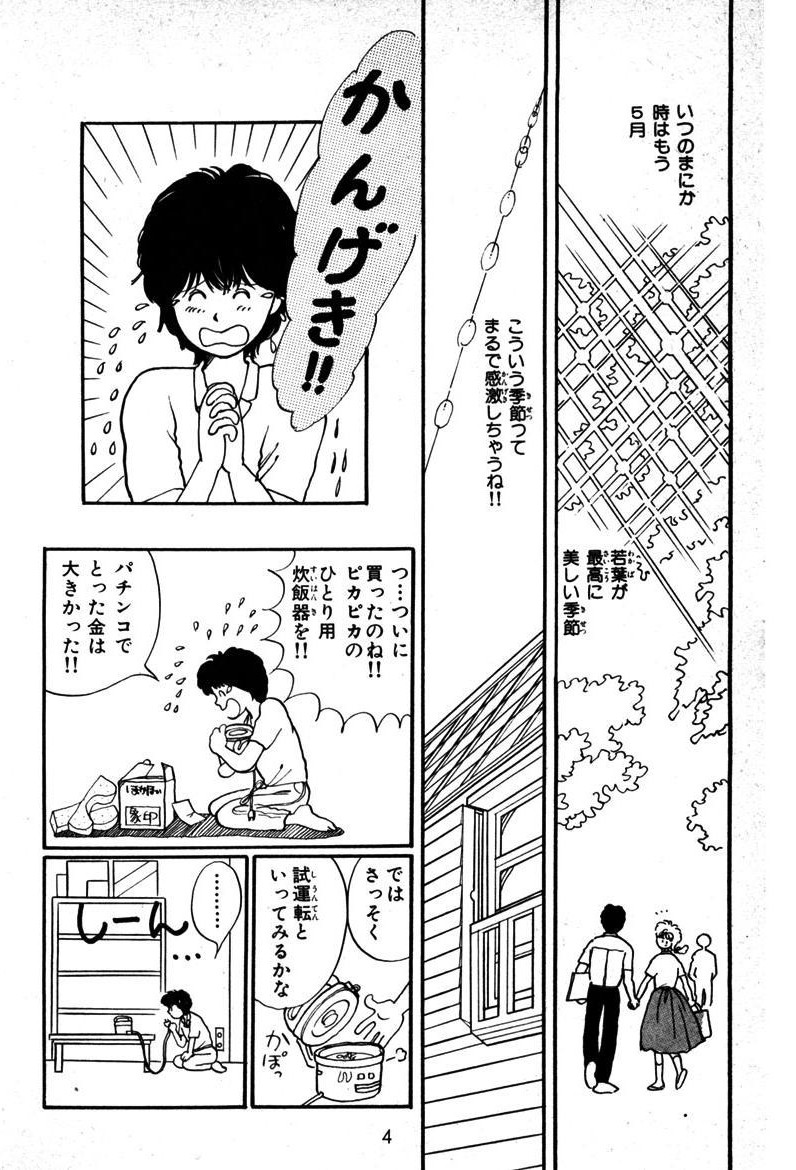}
    \caption{Excerpt of panel sequence from the story arc.}
    \label{fig:sequence-panels}
  \end{subfigure}
  \hfill
  \begin{subfigure}[t]{0.6\linewidth}
    \centering
    \includegraphics[width=\linewidth]{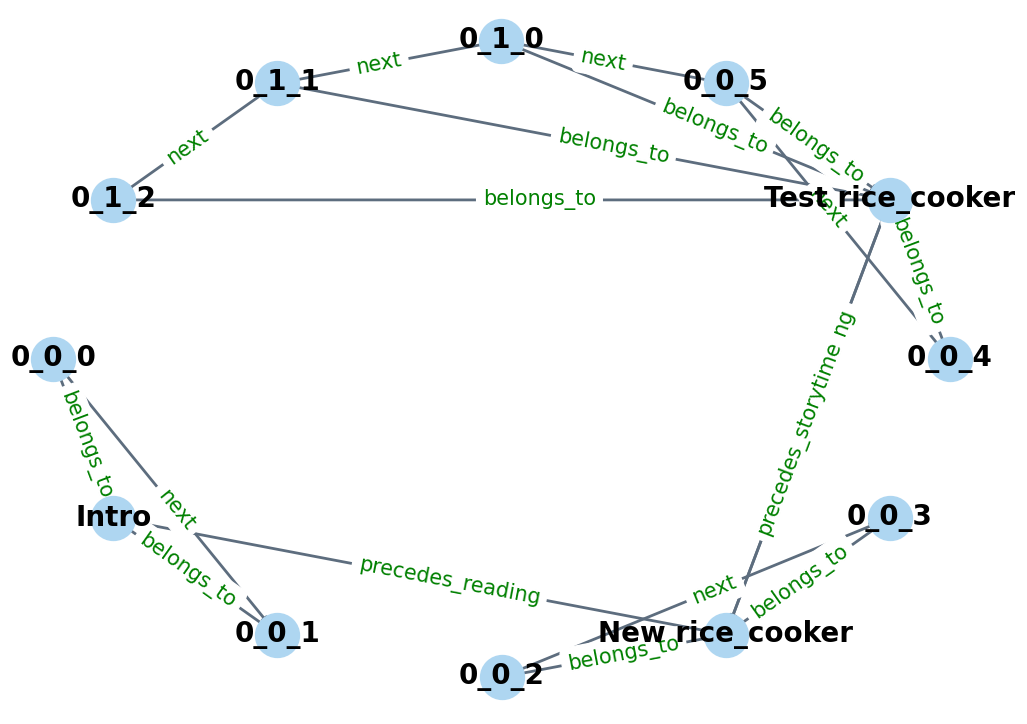}
    \caption{Corresponding sequence-level temporal knowledge graph.}
    \label{fig:sequence-graph}
  \end{subfigure}
  \caption{Sequence-level illustration linking visual panels to their temporal knowledge graph. 
  (a) shows the source panel sequence, while (b) depicts narrative relations such as reading order and event segmentation.}
  \label{fig:sequence-kg}
\end{figure}

Figure~\ref{fig:sequence-kg} extends this perspective to the sequence level. Subfigure~(a) presents a short panel sequence, while subfigure~(b) shows the temporal knowledge graph that links panels and event segments through directed edges encoding both reading order and narrative-time progression. This enables timeline reconstruction and supports analysis of narrative phenomena such as flashbacks, ellipses, or non-linear progression.

\begin{figure}[t]
  \centering
  \includegraphics[width=\linewidth]{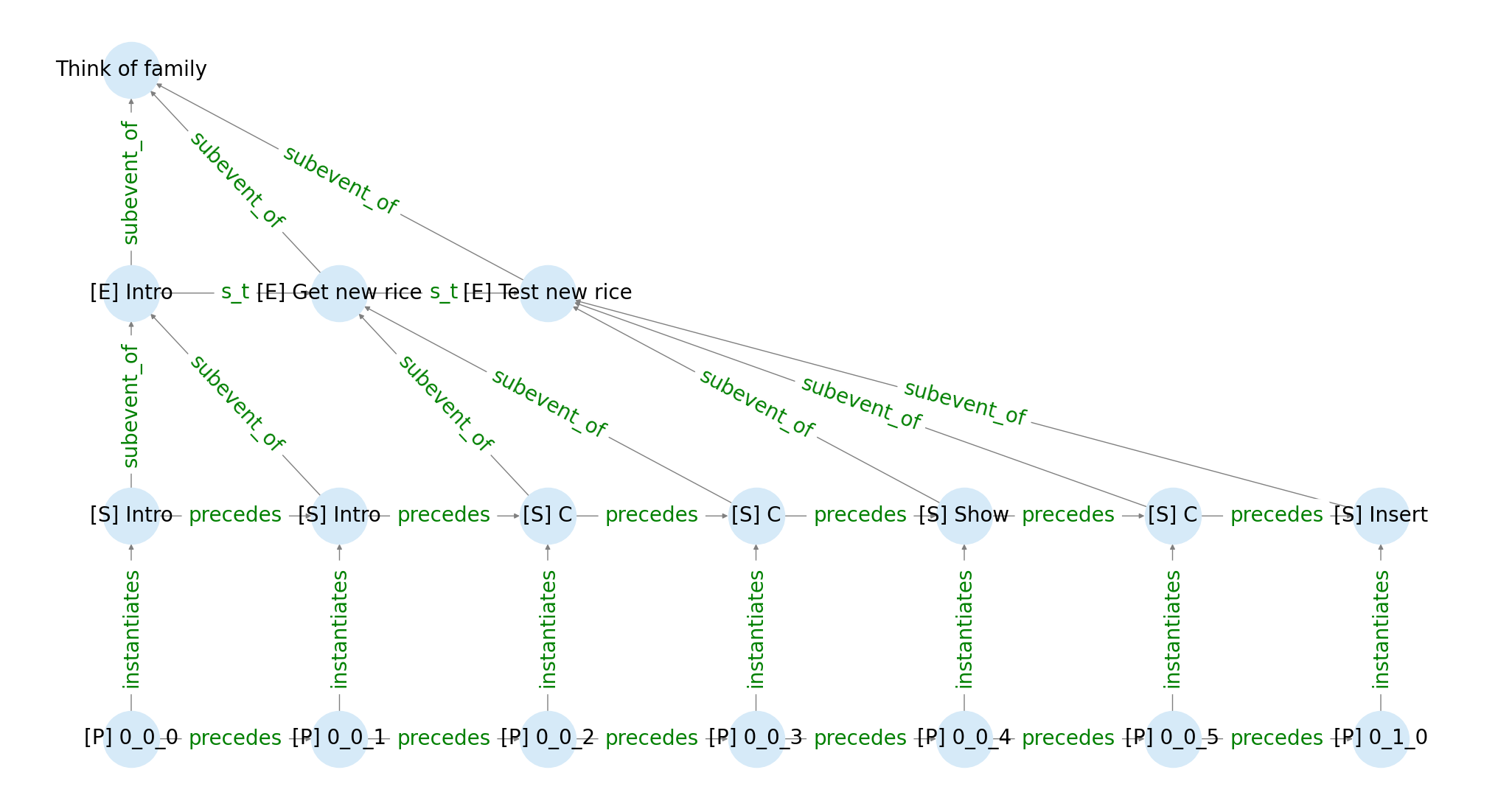}
  % \caption{Partial event-level semantic knowledge graph focused on the reduced macro-event \textit{Think of family} for visibility. This subset includes 18 nodes across four layers, macro-event, events, event segments, and panels, connected by structured edges (\texttt{subevent\_of}, \texttt{precedes\_reading}, \texttt{instantiates}). The layout supports hierarchical reasoning and narrative structure interpretation.}
    \caption{Reduced event-level semantic knowledge graph centered on the macro-event \textit{Think of family}. The visualization shows a simplified subset for readability, including nodes across four layers: macro-event, events, event segments, and panels—connected by structured edges (\texttt{subevent\_of}, \texttt{precedes\_reading}, \texttt{story\_time}, \texttt{instantiates}). 
    This layout highlights the hierarchical organization of narrative structure and supports interpretable reasoning across levels.}
      
  \label{fig:event-kg}
\end{figure}

At a higher level of abstraction, Figure~\ref{fig:event-kg} illustrates part of the event hierarchy focused on the macro-event \textit{Think of family}. The graph spans four layers—macro-event, event, event segment, and panel, connected via relations such as \texttt{subevent\_of}, \texttt{precedes\_reading}, and \texttt{instantiates}. This structure supports queries about event comparison, causal ordering, and narrative summarization.

Together, these qualitative examples demonstrate how the framework represents narrative semantics across multiple levels. By linking panel content to temporal flow and event abstraction, the hierarchical graphs provide modular, interpretable scaffolds for cross-level reasoning tasks.

% \begin{figure}[t]
%   \centering
%   \includegraphics[width=0.9\linewidth]{Images/panel_0_0_4.png}
%   \caption{Panel-level knowledge graph representing multimodal content within a single panel. Nodes correspond to visual and textual entities; edges encode actions, semantic roles, and dialogue associations. This representation supports localized reasoning over character behavior, speech attribution, and visual-textual alignment.}
%   \label{fig:panel-kg}
% \end{figure}

% \begin{figure}[t]
%   \centering
%   \includegraphics[width=0.9\linewidth]{Images/Sequence_example.png}
%   \caption{Sequence-level temporal knowledge graph capturing narrative, time, and reading order relations across panels and event segments. Directed edges encode sequencing, enabling timeline reconstruction and interpretable narrative flow analysis.}
%   \label{fig:sequence-kg}
% \end{figure}

\subsection{Quantitative Evaluation}

To assess the framework’s symbolic reasoning capabilities, we evaluate four representative tasks using graph traversal against human-annotated ground truth. The tasks include action retrieval, dialogue tracing, character appearance mapping, and panel timeline reconstruction. Each task is aligned with an interpretability-oriented evaluation focus (e.g., action recovery, dialogue recall), rather than predictive accuracy. We report coverage—the proportion of annotated content captured by the graph traversal—alongside token-level F1-scores that measure structural and semantic alignment. The results are summarized in Table~\ref{tab:reasoning-results}.

\begin{table}[t]
\centering
\scriptsize
\caption{Performance across four symbolic reasoning tasks using graph-based traversal over the hierarchical knowledge graph. Each task is motivated by an interpretability-oriented evaluation focus, with results reported separately for the two annotated story arcs.}
\label{tab:reasoning-results}
\begin{tabular}{llcccc}
\hline
\multirow{2}{*}{\textbf{Task}} & \multirow{2}{*}{\textbf{Focus}} & \multicolumn{2}{c}{\textbf{Book 0}} & \multicolumn{2}{c}{\textbf{Book 1}} \\
\cline{3-6}
 & & \textbf{Cov. \%} & \textbf{F1} & \textbf{Cov.\%} & \textbf{F1} \\
\hline
Action retrieval             & Action recovery   & 97.3 & 0.96 & 98.0 & 0.97 \\
Dialogue tracing             & Dialogue recall   & 100  & 1.00 & 100  & 0.93 \\
Character appearance mapping & Entity recall     & 100  & 1.00 & 100  & 1.00 \\
Timeline reconstruction      & Seq. ordering     & 100  & 1.00 & 100  & 0.96 \\
\hline
\end{tabular}
\end{table}

As shown in Table~\ref{tab:reasoning-results}, all four tasks achieved near-perfect coverage, confirming that the annotated graphs consistently encode the necessary narrative information. F1-scores range from 0.93 to 1.00, with small variations across tasks. For example, the lower score in dialogue tracing (0.93 in Book 1) reflects prototype-level constraints such as reliance on lexical string-matching, rather than a weakness of the underlying framework. In principle, the same schema can support alternative implementations, including semantic normalization or hybrid neural–symbolic retrieval methods, which would address these prototype-specific limitations.

\subsection{Results and Interpretation}

The system achieves high performance across all evaluated tasks. Dialogue tracing, character mapping, and panel sequence inference attain perfect F1 scores, indicating full alignment with annotated references. These results confirm the internal coherence of the knowledge graph and its effectiveness for cross-level symbolic reasoning.

Action retrieval yields an F1 score of 0.96. The small discrepancy is attributed to lexical variation in action annotations (e.g., \texttt{insert} vs. \texttt{insert\_into}) that are semantically equivalent but not normalized in comparison. Incorporating semantic similarity metrics or synonym handling may further improve accuracy in future work.

In summary, the framework demonstrates that structured symbolic representation can effectively support narrative reasoning tasks across visual and textual modalities. These results validate the proposed approach as an interpretable and extensible foundation for narrative modeling and analysis in multimodal storytelling domains.

\section{Discussion}

This section reflects on the experimental outcomes and broader implications of the proposed hierarchical knowledge graph framework. We examine the framework’s reasoning performance, scalability, and practical utility while also identifying key limitations and directions for future work. These findings are contextualized within larger goals in structured narrative modeling, symbolic reasoning, and multimodal storytelling.

\subsection{Performance and Results}

The evaluation results confirm that the proposed framework supports interpretable and accurate reasoning across multiple levels of narrative abstraction. Perfect F1 scores in dialogue trace, character appearance mapping, and panel timeline reconstruction suggest that the graph encodes consistent structural and semantic relationships across modalities. These findings demonstrate that symbolic representations, when coupled with hierarchical event segmentation, provide a strong foundation for narrative-aligned inference.

The slightly reduced F1 score (0.96) observed in the action retrieval task reflects challenges in handling lexical variation within verb labels. Performance discrepancies typically arise from semantically similar but lexically distinct forms (e.g., \texttt{insert} vs. \texttt{insert\_into}, \texttt{give} vs. \texttt{hand\_over}) being treated as mismatches in token-level comparisons. Addressing this limitation may involve incorporating semantic alignment methods such as verb clustering, synonym expansion, or ontology-based normalization. These enhancements could improve the robustness of the system without sacrificing interpretability.

Overall, the results validate the framework’s ability to support both fine-grained and high-level narrative reasoning, reinforcing its potential in a wide range of visual storytelling applications.

% \subsection{Scalability and Extensibility}

% The proposed framework is designed with modular scalability in mind. Each narrative unit, whether a panel, segment, event, or macro-event, is modeled as a discrete node, enabling incremental graph construction without reconfiguring existing structures. As additional content is introduced, new cross-level relations (e.g., \textit{instantiates}, \textit{subevent-of}) can be dynamically added to support longer or more complex narratives.

% Despite this architectural flexibility, the annotation process remains a bottleneck. Manual labeling of entities, actions, and narrative groupings requires substantial human effort, particularly at higher levels of abstraction. Future development will focus on model-assisted annotation pipelines that incorporate pre-annotation using vision-language models and support interactive refinement. Such tools can significantly reduce annotation costs while preserving semantic fidelity.

% Although our experiments focused on two annotated arcs, the framework is designed to generalize to longer or more varied narratives. The modular event hierarchy and graph schema do not depend on specific artistic style, making them adaptable to other comics, graphic novels, or visual storytelling media.
\subsection{Scalability and Extensibility}

The proposed framework is designed with modular scalability in mind. Each narrative unit, whether a panel, segment, event, or macro-event, is modeled as a discrete node, enabling incremental graph construction without reconfiguring existing structures. As additional content is introduced, new cross-level relations (e.g., \textit{instantiates}, \textit{subevent-of}) can be dynamically added to support longer or more complex narratives.

Despite this architectural flexibility, the annotation process remains a bottleneck. Manual labeling of entities, actions, and narrative groupings requires substantial human effort, particularly at higher levels of abstraction. Future development will focus on model-assisted pipelines that incorporate pre-annotation using vision–language models and support interactive refinement. Such tools can significantly reduce annotation costs while preserving semantic fidelity.

Although our experiments concentrated on two annotated arcs, the framework itself is not limited to these cases. The event hierarchy and graph schema are independent of artistic style or genre, and thus can extend to other comics, graphic novels, and multimodal storytelling media. In principle, the same schema can also be applied to longer narratives or adapted to interactive formats, making the framework broadly extensible beyond the current evaluation.

\subsection{Strengths of the Proposed Framework}

A central strength of this framework lies in its unified representation of multimodal narrative structure. By integrating visual, textual, and structural information across panel-level, temporal, and event hierarchies, the system supports interpretable reasoning across multiple levels of abstraction. This capability enables both localized content analysis (e.g., character actions) and global story comprehension (e.g., event progression and structure alignment).

Moreover, the symbolic nature of the graph provides transparency that is often lacking in end-to-end neural systems. Each reasoning step is encoded explicitly, allowing users to trace how specific outputs are derived from the underlying structure. This level of explainability is especially beneficial in applications such as narrative analysis, educational media, and authoring tools—contexts where interpretability and traceability are essential.

\subsection{Limitations and Challenges}

Despite its strengths, the framework has several limitations. First, the current evaluation relies on exact lexical string-matching to measure overlap between annotated and retrieved outputs. This choice simplifies implementation but underestimates semantic equivalence, since paraphrases or minor variations are counted as mismatches. Future work should incorporate semantic similarity models, structured ontologies, or embedding-based matching to provide more expressive evaluation.

Second, although the graph architecture itself scales well, producing high-quality annotations is time-intensive and requires narrative expertise. Macro-event and event-level labeling involve abstract judgments that are difficult to automate. Semi-supervised pipelines, controlled natural language templates, and narrative structure prediction models could help reduce this annotation burden while preserving quality.

Finally, it is important to distinguish between framework-level design and prototype-level constraints. The hierarchical schema and cross-level event relations are general features of the framework and remain valid regardless of implementation. By contrast, current limitations, such as reliance on JSON-based traversal and string-matching evaluation, reflect choices made in the present prototype. These can be replaced with more advanced semantic matching techniques or hybrid neural–symbolic methods in future iterations.

% \subsection{Applications and Future Work}

% The proposed framework supports a broad spectrum of applications in narrative generation, analysis, and interactive media. As a structured planning layer, the graph can facilitate automated storytelling by organizing character actions, temporal sequences, and dialogue flows into coherent arcs. In interactive contexts such as games or comics, the graph enables dynamic querying, branching narratives, and author-guided scene planning.

% Analytically, the framework offers tools for narrative summarization, structure alignment, and semantic retrieval. For example, it can compress long storylines into abstract summaries, compare plot variations across media versions, or support semantic search over symbolic narrative elements.

% Future directions include extending the framework to handle longer and more diverse narratives, including animated or non-linear formats such as interactive fiction. Automating parts of the annotation process using pretrained models will be key to scaling deployment. We also plan to develop authoring interfaces that allow creators to interact with, edit, and reason over the graph structure directly, enabling collaborative storytelling and human-AI co-creation.

% These directions aim to extend the framework’s impact across narrative-centric AI applications, bridging symbolic representation with multimodal content and interactive storytelling.

\subsection{Applications and Future Work}

The proposed framework has potential applications in narrative analysis, comprehension, and eventually generation and interactive media. As a structured planning layer, the graph can organize character actions, temporal sequences, and dialogue flows into coherent arcs, providing a foundation for automated storytelling systems. In interactive contexts such as games or comics, the graph structure can support dynamic querying, branching narratives, and author-guided scene planning.

Analytically, the framework offers tools for narrative summarization, structure alignment, and semantic retrieval. It can be used to compress long storylines into abstract summaries, compare plot variations across different media versions, or support semantic search over symbolic narrative elements.

Future directions include extending the framework to longer and more diverse narratives, including animated or non-linear formats such as interactive fiction. Automating parts of the annotation process with pretrained models will be essential for scaling. In addition, the schema may be extended to authoring interfaces that allow creators to interact with, edit, and query the graph structure directly, enabling collaborative storytelling and human–AI co-creation.

Together, these applications and extensions highlight the framework’s capacity to bridge symbolic representation with multimodal content and lay the groundwork for future interactive storytelling systems.

% \section{Future Work}
% \input{7_Future}
\section{Conclusion}

This paper presents a hierarchical knowledge graph framework for visual narrative comprehension, using comics as a representative domain for multimodal storytelling. 
The proposed approach encodes panels, events, and macro-events as structured graph representations that capture temporal, semantic, and multimodal relationships. 
This architecture enables interpretable reasoning across multiple levels of narrative abstraction.

We evaluate the framework through four symbolic reasoning tasks: action retrieval, dialogue tracing, character appearance mapping, and panel timeline reconstruction. 
Across these tasks, the framework shows consistently high alignment with human annotations, demonstrating the effectiveness of structured graph-based reasoning in supporting both fine-grained and abstract narrative understanding in visually grounded media.

Beyond the reported evaluation, the framework offers a modular and extensible foundation for narrative-centered AI systems. Its symbolic design supports transparency, traceability, and hierarchical traversal, features that are valuable for applications requiring interpretable narrative representations. While the present study focuses exclusively on comprehension, the same schema could inform future extensions to automated storytelling, interactive systems, and educational applications.

Future directions include extending the framework to longer and more diverse narrative forms, integrating semantic normalization techniques, and supporting scalable annotation via model-assisted pipelines. We also envision investigating author-facing tools that leverage the framework to interactively build, edit, and query narrative graphs, enabling new forms of collaborative and explainable storytelling. Taken together, this work lays the groundwork for more interpretable, adaptable, and semantically expressive narrative intelligence systems.

\section*{Data and Code Availability}
All annotated knowledge graphs, annotation schema, and implementation scripts are publicly available at:  
\begin{itemize}
    \item \textbf{Hierarchical framework and reasoning scripts:} \url{https://github.com/RimiChen/2025_VN_Modeling}
    \item \textbf{Semantic normalization extension:} \url{https://github.com/RimiChen/2025_VN_Semantic_Normalization}
\end{itemize}

The repositories include hierarchical annotations, graph construction pipelines, reasoning functions, visualization tools, 
and a semantic normalization variant. These resources support transparency, reproducibility, and future extensions of the framework.

\bibliographystyle{splncs04}
\bibliography{reference}




% LNCS Homepage, \url{http://www.springer.com/lncs}, last accessed 2023/10/25
% \end{thebibliography}
\end{document}